\documentclass{emulateapj}

\usepackage{graphicx}
\usepackage{amssymb,amsfonts,amsmath,amstext,amsgen,amsopn,amsxtra,indentfirst,lscape,times,rotating}

\shorttitle{Stability of Super Earth Atmospheres}
\shortauthors{Heng  \& Kopparla}

\begin{document}

\title{On the Stability of Super Earth Atmospheres}

\author{Kevin Heng\altaffilmark{1,3}}
\author{Pushkar Kopparla\altaffilmark{2}}

\altaffiltext{1}{ETH Z\"{u}rich, Institute for Astronomy, Wolfgang-Pauli-Strasse 27, CH-8093, Z\"{u}rich, Switzerland}
\altaffiltext{2}{ETH Z\"{u}rich, Institute for Atmospheric and Climate Science, Universit\"{a}tstrasse 16, CH-8092, Z\"{u}rich, Switzerland}
\altaffiltext{3}{Zwicky Prize Fellow}

\begin{abstract}
We investigate the stability of super Earth atmospheres around M stars using a 7-parameter, analytical framework.  We construct stability diagrams in the parameter space of exoplanetary radius versus semi-major axis and elucidate the regions in which the atmospheres are stable against the condensation of their major constituents, out of the gas phase, on their permanent nightside hemispheres.  We find that super Earth atmospheres which are nitrogen-dominated (``Earth-like") occupy a smaller region of allowed parameter space, compared to hydrogen-dominated atmospheres, because of the dual effects of diminished advection and enhanced radiative cooling.  Furthermore, some super Earths which reside within the habitable zones of M stars may not possess stable atmospheres, depending on the mean molecular weight and infrared photospheric pressure of their atmospheres.  We apply our stability diagrams to GJ 436b and GJ 1214b, and demonstrate that atmospheric compositions with high mean molecular weights are disfavoured if these exoplanets possess solid surfaces and shallow atmospheres.  Finally, we construct stability diagrams tailored to the \textit{Kepler} dataset, for G and K stars, and predict that about half of the exoplanet candidates are expected to habour stable atmospheres if Earth-like conditions are assumed.  We include 55 Cancri e and CoRoT-7b in our stability diagram for G stars.
\end{abstract}


\section{Introduction}

The lower effective temperatures of M stars (or red dwarfs)---which comprise about three-quarters of the stellar population---render the detection of orbiting, close-in, Earth-like exoplanets amenable to both the transit and radial velocity techniques \citep{char09}.  The immense interest in hunting for exoplanets around these stars stems from the fact that the habitable zone is located at $\sim 0.01$--0.1 AU, rather than $\sim 1$ AU, away from these stars \citep{tarter07}.  A consequence of the close proximity is that the exoplanets are expected to be tidally locked (or spin synchronized),\footnote{Strictly speaking, the terms ``tidally locked" and ``spin synchronized" are only synonymous for an exoplanet residing on a circular orbit \citep{heller11}.  We use these terms interchangeably as we do not account for the effects of an eccentric orbit.} meaning that they possess permanent dayside and nightside hemispheres.  This expectation has led to theoretical concerns that the atmospheres may undergo collapse, because if heat is not redistributed efficiently from the dayside to the nightside it may lead to the dominant chemical species condensing out on the nightside (e.g., \citealt{j97,j03}).  For example, \cite{wordsworth11} constructed three-dimensional, general circulation models tailored to the study of Gliese 581d and elucidated scenarios for atmospheric stability.  Generally, the discovery of close-in super Earths (e.g., \citealt{mayor09,char09b,vogt10,ang12,delfosse12}) has inspired a number of theoretical studies of their possible atmospheres (e.g., \citealt{ms10,rs10,cm11,hmp11,hv11,hfp11,p11,valencia11,menou12}), which either focus on specific case studies or explore a limited range of parameter space.

Specifically, we consider an exoplanet possessing a solid surface which is enveloped by a shallow atmosphere.  Whether the surface is made of rock, ice, etc, is unimportant, insofar as it provides a solid, lower boundary for the atmosphere.  The atmosphere is assumed to be shallow enough that it may be characterized by advective and radiative time scales which are constant over its entire depth.  When the atmosphere is predominantly radiative, heat is thermally re-emitted before it is able to be advected to the permanent nightside.  The nightside becomes arbitrarily cold,\footnote{But see discussion of caveats in \S\ref{subsect:caveats}.} thus leading to the condensation of its constituents from the gas to the liquid or solid phases.  When this situation occurs, the atmosphere is unlikely to survive for a duration comparable to the stellar age and we term the atmosphere to have ``collapsed".  This situation is distinct from the phenomenon of atmospheric escape (e.g., \citealt{yelle04,mc09}), which concerns the (often slow) vertical flow of material out of the gravitational potential of the exoplanet.  When the atmosphere is predominantly advective, the efficient redistribution of heat to the nightside provides a necessary but insufficient condition for the nightside atmosphere to exist in its gaseous form---whether it does depends on the thermodynamics of phase changes of the chemical species being considered, which provides the sufficient condition.  If the atmosphere remains in the gas phase, then we term it to be ``stable".  In other words, an atmosphere which is sufficiently advective \emph{and} warm will survive for a long time.

In the present paper, the main question we are addressing is: \emph{what is the simplest model one can construct to broadly understand the stability of super Earth atmospheres?}  To this end, we introduce an analytical framework which allows for an efficient, broad exploration of parameter space, while keeping the number of free parameters involved to a minimum.  The key outcome of our study is the construction of a stability diagram, which allows one to judge if a super Earth atmosphere is likely to survive for a duration comparable to the age of its host star.  In \S\ref{sect:method}, we describe our methodology.  In \S\ref{sect:stability}, we present the basic stability diagram as well as variations of it applied to GJ 436b and GJ 1214b.  In \S\ref{sect:discussion}, we discuss the caveats associated with our model, the implications of our results and also apply our stability diagrams to the \textit{Kepler} dataset of exoplanets and exoplanetary candidates, as well as 55 Cancri e and CoRoT-7b.

\section{Methodology}
\label{sect:method}

To efficiently explore a wide range of parameter space, we construct a 7-parameter model with the following inputs:
\begin{enumerate}

\item The spatial separation between the exoplanet and its host star ($a$);

\item The radius of the exoplanet ($R$);

\item The bulk mass density of the exoplanet ($\rho_0$);

\item The pressure level associated with the infrared photosphere ($P$);

\item The mean molecular weight of atmospheric atoms/molecules ($\mu$);

\item The number of degrees of freedom of the atmospheric gas ($n_{\rm dof}$);

\item The Bond albedo of the exoplanetary atmosphere (${\cal A}$).

\end{enumerate}
These basic parameters yield all of the other secondary parameters.  We assume our exoplanets to be spherical, such that the mass is described by $M = 4\pi \rho_0 R^3/3$.  The surface gravity is given by $g = 4\pi G \rho_0 R/3$, where $G$ is the universal gravitational constant.  In addition, the stellar mass ($M_\star$), radius ($R_\star$) and effective temperature ($T_\star$) need to be specified.

\subsection{Stellar Irradiation and Orbital Parameters}

The incident flux impinging upon the substellar point is 
\begin{equation}
{\cal F}_0 = \sigma_{\rm SB} T^4_{\rm irr},
\end{equation}
where $\sigma_{\rm SB}$ is the Stefan-Boltzmann constant and the irradiation temperature is
\begin{equation}
T_{\rm irr} = T_\star \left( \frac{R_\star}{a} \right)^{1/2} \left( 1 - {\cal A} \right)^{1/4}.
\end{equation}
If one adopts parameter values appropriate to Earth (and ${\cal A}=0$), one obtains the solar constant, ${\cal F}_0 \approx 1370$ W m$^{-2}$.  

Since the assumption of tidal locking is made, the orbital and rotational frequency are equal and given by (assuming that $M_\star \gg M$)
\begin{equation}
\Omega = \left( \frac{GM_\star}{a^3} \right)^{1/2}.
\end{equation}
It is thus clear that $T_{\rm irr}$ (or ${\cal F}_0$) and $\Omega$ need to be varied self-consistently as one changes $a$.

\subsection{Thermodynamics}
\label{subsect:thermo}

\begin{table*}
\centering
%\begin{minipage}{140mm}
\caption{Table of Thermodynamic Quantities for Various Molecular Species}
\label{tab:condensation}
{\footnotesize
\begin{tabular}{lcccccc}
\hline\hline
\multicolumn{1}{c}{Species} & \multicolumn{1}{c}{$\mu$} & \multicolumn{1}{c}{$L_{\rm avg}$ (J kg$^{-1}$)}  & \multicolumn{1}{c}{$T_{\rm con,0}$ (K)} & \multicolumn{1}{c}{$T_{\rm crit}$ (K)} & \multicolumn{1}{c}{$P_{\rm sat,0}$ (bar)} & \multicolumn{1}{c}{$P_{\rm crit}$ (bar)}\\
\hline
\vspace{2pt}
H$_2$O &  18 & $2.7 \times 10^6$ & 5845 & 647.1 & $1.2 \times 10^7$ & 221 \\
CH$_4$ & 16 & $5.7 \times 10^5$ & 1097 & 190.44 & $2.1 \times 10^4$ & 45.96 \\
CO$_2$ & 44 & $5.0 \times 10^5$ & 2646 & 304.2 & $1.1 \times 10^6$ & 73.825 \\
N$_2$ & 28 & $2.3 \times 10^5$ & 775 & 126.2 & $2.7 \times 10^4$ & 34.0 \\
O$_2$ & 32 & $2.5 \times 10^5$ & 962 & 154.54 & $7.4 \times 10^4$ & 50.43 \\
NH$_3$ & 17 & $1.8 \times 10^6$ & 3680 & 405.5 & $9.2 \times 10^6$ & 112.8 \\
H$_2$ & 2 & $4.5 \times 10^5$ & 108 & 33.2 & $1.7 \times 10^2$ & 12.98\\
\hline
\hline
\end{tabular}}\\
%\end{minipage}
\end{table*}

\begin{figure}
\centering
\includegraphics[width=\columnwidth]{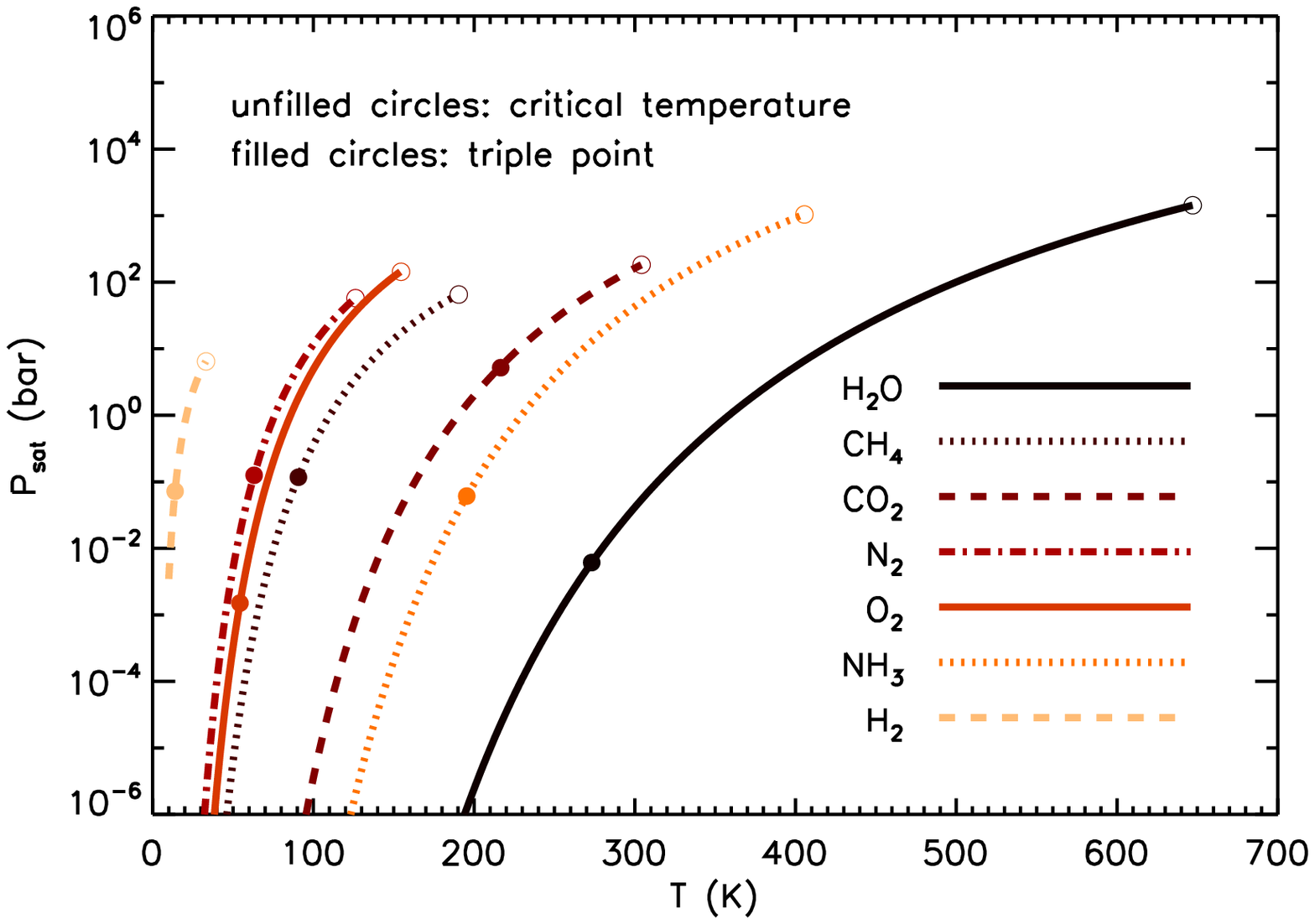}
\includegraphics[width=\columnwidth]{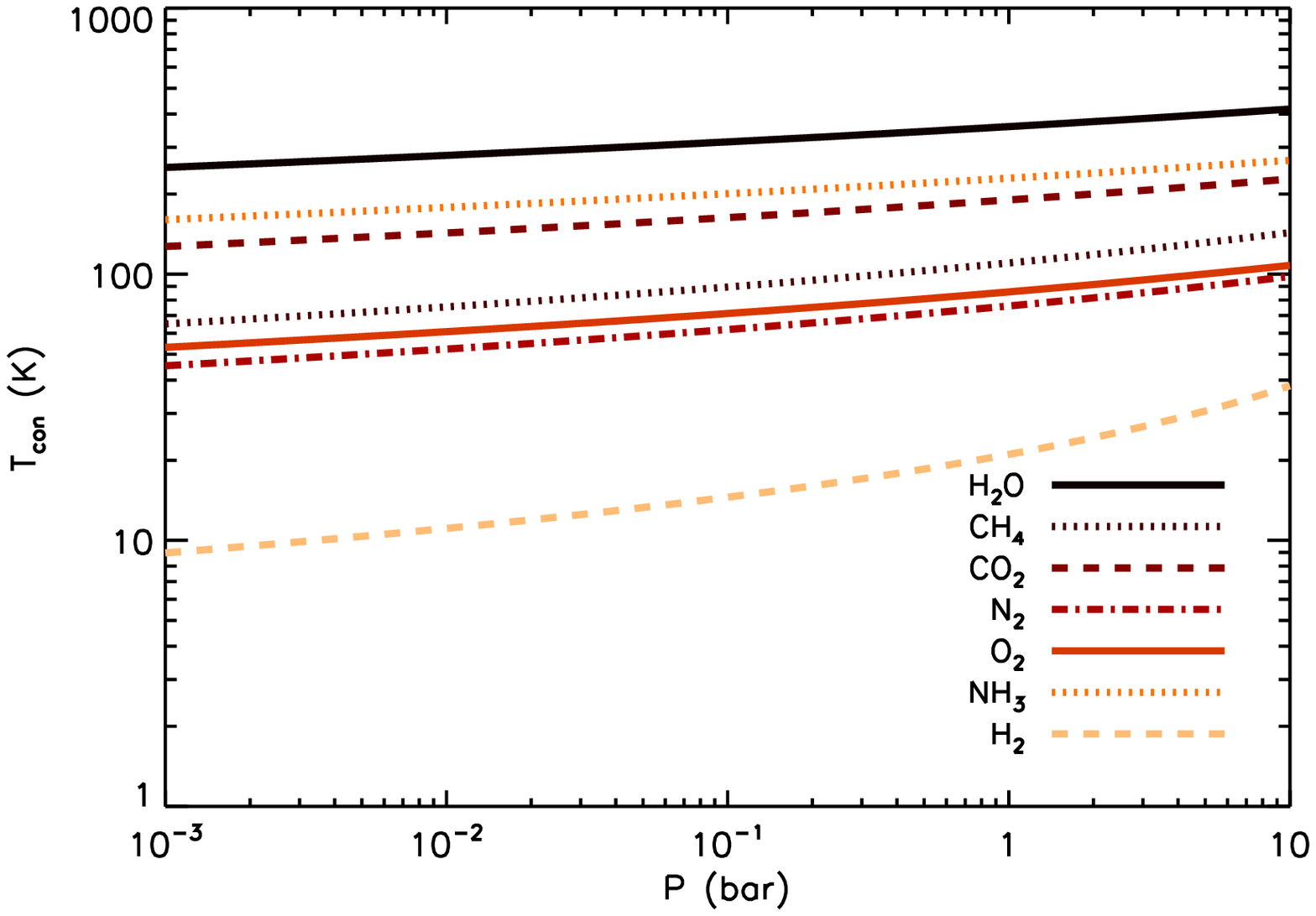}
\caption{Top panel: approximate phase diagrams of various molecular species as computed using the Clausius-Clapeyron equation.  At a given temperature, chemical species existing at a pressure above their respective saturation pressure curves are in the liquid or solid phase.  Bottom panel: condensation temperature as a function of pressure.  At a given pressure, chemical species existing at a temperature above $T_{\rm con}$ are in the gas phase.}
\label{fig:condensation}
\end{figure}

The mean molecular mass $m = \mu m_{\rm H}$ (with $m_{\rm H}$ denoting the mass of a hydrogen atom) and the number of degrees of freedom of the atmospheric gas $n_{\rm dof}$ collectively describe the thermodynamics and are manifested macroscopically via the adiabatic gas index,
\begin{equation}
\Gamma = 1 + \frac{2}{n_{\rm dof}},
\end{equation}
as well as the adiabatic coefficient \citep{pierrehumbert10},
\begin{equation}
\kappa = \frac{2}{2 + n_{\rm dof}} = \frac{{\cal R}}{c_P},
\end{equation}
where ${\cal R} = {\cal R}^\ast/\mu$ is the specific gas constant and ${\cal R}^\ast = 8314.5$ J K$^{-1}$ kg$^{-1}$ is the universal gas constant.  The sound speed is $c_s = ( \Gamma k_{\rm B} T_{\rm irr}/ m )^{1/2}$, where $k_{\rm B}$ is the Boltzmann constant.  Furthermore, the specific heat capacity,
\begin{equation}
c_P = \frac{\left(2 + n_{\rm dof} \right) {\cal R}^\ast}{2\mu},
\end{equation}
is used in evaluating the radiative time scale $t_{\rm rad}$. 

As an example, atmospheres dominated by molecular hydrogen have $\mu=2$ and $n_{\rm dof}=5$, such that $\Gamma = 7/5$, $\kappa=2/7$ and ${\cal R} = 4157.25$ J K$^{-1}$ kg$^{-1}$.  Dry, terrestrial air (dominated by nitrogen and oxygen) has a mean molecular weight of $\mu \approx 28.97$, such that ${\cal R} \approx 287$ J K$^{-1}$ kg$^{-1}$.  As a further example, we note that the model with $\mu \approx 17$ from \cite{menou12} is termed ``Water", with relevance to a possible subclass of H$_2$O-dominated super Earths known as ``water worlds" \citep{char09b,rs10,bean11}.  In practice, the effective number of degrees of freedom $n_{\rm dof}$ is not an integer (e.g., see Table 2.1, pg. 92, of \citealt{pierrehumbert10}).  For example, CO$_2$ has $n_{\rm dof} \approx 6.8$, O$_2$ has $n_{\rm dof} \approx 5.1$ and H$_2$ has $n_{\rm dof} \approx 5.2$.  For simplicity and to not obscure the salient physics of the model, we adopt $n_{\rm dof}=5$.

To determine if a given chemical species exists in the gas or liquid/solid phase, one needs to have knowledge of the saturation pressure $P_{\rm sat}$.  If the atmospheric pressure exceeds the saturation pressure, then the molecule (or atom) will condense out of the gas phase and into the liquid/solid phase.  The dependence of $P_{\rm sat}$ on the temperature $T$ is governed by the Clausius-Clapeyron equation (e.g., \S2.6 of \citealt{pierrehumbert10}),
\begin{equation}
\frac{dP_{\rm sat}}{dT} = \frac{P_{\rm sat} T_{\rm con,0}}{T^2},
\end{equation}
where $T_{\rm con,0} \equiv L_{\rm avg}/{\cal R}$ is the characteristic condensation temperature.  Depending on whether one is describing the gas-liquid or gas-solid phase change, one needs to specify the latent heat of condensation or sublimation, respectively.  For most gases, the difference between these two latent heats is small ($\approx 5$--10\%); an exception is CO$_2$, where the difference is about 33\%.  For simplicity, we adopt the average value of these two latent heats ($L_{\rm avg}$) using Table 2.1 of \cite{pierrehumbert10}.  Furthermore, the value of $L_{\rm avg}$ remains fairly constant unless $T \gg T_{\rm con,0}$.  Thus, the Clausius-Clapeyron equation can be integrated to obtain
\begin{equation}
P_{\rm sat} = P_{\rm sat,0} ~\exp{\left(- \frac{T_{\rm con,0}}{T} \right)},
\label{eq:psat}
\end{equation}
where the normalization constant $P_{\rm sat,0}$ is computed using the pressure and temperature associated with the triple point of a given chemical species.  In Table \ref{tab:condensation}, we list $L_{\rm avg}$, $T_{\rm con,0}$ and $P_{\rm sat,0}$ for several molecular species.  The top panel of Figure \ref{fig:condensation} shows the corresponding (and approximate) phase diagrams, which include the triple points associated with each species.  

Our use of the saturation pressure curves is valid as long as one is not above the temperature $T_{\rm crit}$ and the pressure $P_{\rm crit}$ associated with the critical point.  In this regime, the atmosphere exists as a supercritical fluid where distinct liquid and gas phases do not exist.  If the atmosphere is above the critical temperature but below the critical pressure, then it exists as a gas.  Since we typically have $P_{\rm crit} > 10$ bar (which is larger than the pressures we consider), we assume our model atmospheres to be gaseous when $T>T_{\rm crit}$.

At the infrared photosphere (located at a pressure $P$), the atmosphere remains in gaseous form if $P<P_{\rm sat}$, which may be recast as the condition $T>T_{\rm con}$ where
\begin{equation}
T_{\rm con} \equiv \frac{T_{\rm con,0}}{\ln{\left( P_{\rm sat,0}/P \right)}}
\label{eq:tmin}
\end{equation}
is the condensation temperature.  The logarithmic factor $\ln( P_{\rm sat,0}/P  )$ implies that $T_{\rm con}$ has a somewhat gentle dependence on chemical composition.  In the bottom panel of Figure \ref{fig:condensation}, we see that $T_{\rm con} \sim 100$ K.  An exception is molecular hydrogen for which $T_{\rm con} \sim 10$ K over a wide range of photospheric pressures.  Generally, it becomes less difficult to condense out a given chemical species as the pressure increases.

\subsection{Infrared Photospheric Pressure}

We assume a thin atmosphere with a Bond albedo ${\cal A}$, such that a fraction $(1-{\cal A})$ of the stellar flux reaches the surface, which resides at a pressure $P_0$.  The starlight is absorbed and re-emitted in the infrared at a photospheric pressure $P<P_0$.  Furthermore, we assume $P \sim P_0$ such that we may apply a shallow water model to represent the entire atmosphere between $P$ and $P_0$.  The atmosphere is also thin in the sense that the radiative time scale $t_{\rm rad}$ is assumed to be approximately constant throughout, at least near the substellar point.  We regard $P$ as a free parameter which characterizes the thickness of the atmosphere.\footnote{This situation is distinct from that of hot Jupiters, where one parametrizes the optical depths by two broadband opacities \citep{hhps12}.  The radiative time scale is generally not constant for hot Jupiters and varies with depth.  The variation of shortwave scattering (i.e., albedo) shifts the photon deposition depth to different heights, resulting in varying degrees of heat redistribution between the dayside and nightside hemispheres \citep{heng12}.}

While we model our atmospheres as consisting purely of molecular hydrogen or nitrogen in our stability diagrams, the assumption is that they also contain minor amounts of other chemical species which absorb strongly in the infrared but constitute an insignificant fraction of the total mass of the atmosphere.  This enables us to define a ``mean gas" with a mean molecular weight and to compute quantities such as $c_P$.  We further assume that these trace constituents are scarce enough that their partial pressures are miniscule, such that they always remain in the gas phase.  In this case, atmospheric stability is dictated by the main constituent (i.e., H$_2$ or N$_2$) existing in the gaseous regime of the phase diagram.  The typical collisional time between the major and minor constituents of the atmosphere is assumed to be the shortest time scale in the system, such that the atmosphere may be modeled as a single fluid.  We neglect any effects associated with clouds or hazes (e.g., see \citealt{hhps12}).

\subsection{Hydrodynamic Shallow Water Model}
\label{subsect:shallow}

We utilize the linear, analytical, steady-state, two-dimensional version of the shallow water model of \cite{sp11} to model the photosphere of a super Earth.  (See also \citealt{sp10}.)  The shallow water system has traditionally been used to model both the terrestrial atmosphere and ocean \citep{matsuno66,lh68,gill80}, but has been generalized by \cite{sp11} to include radiative forcing and drag within the context of spin-synchronized exoplanets.  (See also \citealt{hs09} and references therein.)  Several adjustments are required to mimic the variation of various atmospheric quantities with the strength of stellar irradiation.

Firstly, we need to relate the mean shallow water height $H$ to a characteristic height in the atmosphere, which we identify as the pressure scale height,
\begin{equation}
H = \frac{k_{\rm B} T_{\rm irr}}{m g}.
\end{equation}
Secondly, we set the thermal forcing term $S_0$ (which has physical units of cm s$^{-1}$) to be $H/t_0$, where $t_0 = (gH)^{-1/4} \beta^{-1/2}$ is the characteristic time scale used to non-dimensionalize the forcing and $\beta = 2\Omega/R$, and also adopt a zonal wave number of 0.5.  In other words, the dimensionless forcing is set to be unity.  The third ingredient is the radiative time scale \citep{gy89}, which is the ratio of the thermal energy content per unit area ($c_P T_{\rm irr} \tilde{m}$ where $\tilde{m} = P/g$ is the column mass in hydrostatic equilibrium) to the flux ${\cal F}_0$,
\begin{equation}
t_{\rm rad} = \frac{c_P P}{\sigma_{\rm SB} g T_{\rm irr}^3},
\label{eq:trad}
\end{equation}
which enters into the momentum equations via a radiative drag term ($-\vec{v}/t_{\rm rad}$).

The fourth---and most uncertain---ingredient is the hydrodynamic drag time scale, which we set to be the rotational/orbital period,
\begin{equation}
t_{\rm drag} = \frac{2\pi}{\Omega}.
\label{eq:drag}
\end{equation}
On Earth, the source of hydrodynamic drag originates from the boundary layer \citep{g94}, which is the transition region from a ``no slip" to a ``free slip" condition between the atmosphere and the terrestrial surface.  In the Held-Suarez benchmark test for Earth, the time scale for Rayleigh drag is set to be one Earth day \citep{hs94}.  Our simple assumption for $t_{\rm drag}$ is plausible and commensurate with the fact that the true source of drag in super Earth atmospheres remains to be identified.  Even the simplest implementation of a boundary layer scheme for Earth \citep{tm86,fhz06} requires the specification of four additional parameters, which is currently unjustified given the dearth of astronomical information on this issue.

\subsection{Basic Trends with Strength of Stellar Irradiation}

\begin{figure}
\centering
\includegraphics[width=\columnwidth]{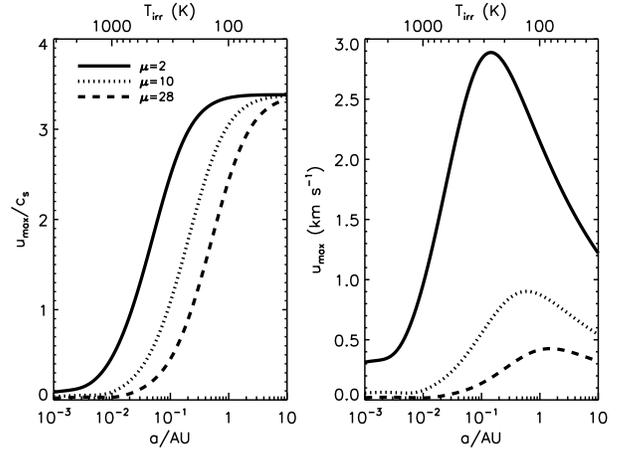}
\caption{Variation of the maximum zonal wind speed, non-dimensionalized (left panel) and in physical units (right panel), with $a$ (or $T_{\rm irr}$), computed using the shallow water model of Showman \& Polvani (2011) which is specialized to tidally-locked exoplanets.  For illustration, we adopt $M_\star = 0.184 M_\odot$, $R_\star = 0.203 R_\odot$ and $T_\star = 3240$ K (appropriate to a M3.5 star) as well as $P=0.1$ bar and ${\cal A}=0$.  Shown also are the curves for different values of the mean molecular weight $\mu$.}
\label{fig:trends}
\end{figure}

To assess the efficiency of heat redistribution from the dayside to the nightside hemisphere of a tidally-locked super Earth, one needs to compare the radiative to the advective time scale \citep{sg02}.  \cite{php12} have shown using three-dimensional simulations of atmospheric circulation---albeit in the context of hot Jupiters---that a reasonable approach to computing the advective time scale is to evaluate
\begin{equation}
t_{\rm adv} \sim \frac{R}{u_{\rm max}},
\end{equation}
where $u_{\rm max} \equiv \mbox{max\{}u\}$ and $u$ is the zonal wind speed.

While we do not expect the shallow water model of \cite{sp11} to reproduce all of the trends obtained from the three-dimensional simulations, we do expect $u_{\rm max}$ to generally increase with $T_{\rm irr}$ when hydrodynamic drag is insignificant.  The right panel of Figure \ref{fig:trends} confirms this expectation for $a \gtrsim 0.1$ AU, a basic trend which is supported by the results obtained from three-dimensional simulations of atmospheric circulation (e.g., \citealt{php12}).  At $a \lesssim 0.1$ AU, the hydrodynamic drag condition we have imposed to mimic the effects of the boundary layer sets in and reverses the trend.  We have checked that if $t_{\rm drag}$ is kept constant (and at a large value), then $u_{\rm max}$ increases monotonically with decreasing $a$.  The left panel of Figure \ref{fig:trends}, which shows the maximum zonal wind speed normalized by the sound speed, illustrates the effects of hydrodynamic drag more succinctly.  

The model also predicts that $u_{\rm max}$ decreases when the mean molecular weight increases, because a larger value of $\mu$ corresponds to stronger radiative forcing (i.e., shorter $t_{\rm rad}$).  Such a trend with $\mu$ is consistent with the simulated results of \cite{menou12}, who---by examining models for GJ 1214b with $\mu \approx 2, 3$ and 17---found that a higher value of $\mu$ leads to a higher ratio of dayside to nightside photospheric flux, implying that heat redistribution becomes less efficient as $\mu$ increases.  We have checked that the qualitative trends in our stability diagrams are similar whether we adopt $t_{\rm drag} = 2\pi/\Omega$ or as a constant, arbitrary value, although the quantitative results do differ.

Three-dimensional numerical solutions of the (more general) hydrodynamic primitive equations (e.g., \citealt{vallis}) predict $\sim1$--2 km s$^{-1}$ zonal flows, which are mildly supersonic, to be present in hypothetical super Earth atmospheres tailored to the study of GJ 1214b \citep{menou12}.  Our estimates for the maximum zonal wind speed are broadly consistent with these calculations.  Supersonic flows are permitted in shallow water models (e.g., \citealt{ant10}), and the assumption of vertical hydrostatic equilibrium does not affect the speed of the horizontally propagating sound waves in general \citep{mac00}.

\subsection{Inefficiency of Thermal Conduction}

We assume that the rocky cores of our model super Earths do not efficiently conduct enough heat from the irradiated dayside hemisphere to the cold nightside hemisphere to prevent atmospheric collapse.  We will now demonstrate that this is a plausible assumption.

In one dimension, the conduction equation reads
\begin{equation}
\frac{\partial T}{\partial t} = \alpha_0 \frac{\partial^2 T}{\partial x^2},
\end{equation}
where $T$ denotes the temperature, $t$ the time, $x$ the spatial coordinate and
\begin{equation}
\alpha_0 = \frac{k_{\rm cond}}{\rho_0 c_{P_0}}
\end{equation}
is the thermal diffusivity of the material of the rocky core.  Here, $c_{P_0}$ denotes the specific heat capacity, at constant pressure, of the rocky core (and not the atmosphere).  Assuming that $T = T(x,t) = T_x(x) T_t(t)$ is a separable function of $x$ and $t$ yields
\begin{equation}
\frac{d T_t}{d t} = - {\cal C} \alpha_0 T_t, 
\end{equation}
which has the solution
\begin{equation}
T_t = T_{t_0} \exp{\left( - {\cal C} \alpha_0 t\right)},
\end{equation}
where $T_{t_0} \equiv T_t(t=0)$ and $-{\cal C}$ is the separation constant.  We identify ${\cal C} = 1/R^2$, such that the characteristic thermal conduction time scale is
\begin{equation}
\begin{split}
t_{\rm cond} =& \frac{\rho_0 c_{P_0} R^2}{k_{\rm cond}} \\
\approx& 4 \times 10^7 \mbox{ yr} ~\left( \frac{\rho_0}{3 \mbox{ g cm}^{-3}} \frac{c_{P_0}}{10^7 \mbox{ erg K}^{-1} \mbox{ g}^{-1}} \right) \\
&\times \left( \frac{R}{R_\oplus} \right)^2 \left( \frac{k_{\rm cond}}{10^{10} \mbox{ erg K}^{-1} \mbox{ s}^{-1}} \right)^{-1}.
\end{split}
\end{equation}
Rocky material is expected to have a thermal conductivity of $k_{\rm cond} \sim 1$ W mK$^{-1} = 10^{10}$ erg K$^{-1}$ s$^{-1}$ and a specific heat capacity of $c_{P_0} \sim 10^7$ erg K$^{-1}$ g$^{-1}$.  Demanding that the radiative cooling time is less than the thermal conduction time yields the condition
\begin{equation}
\begin{split}
T &> \left( \frac{k_{\rm cond}}{\sigma_{\rm SB} R} \right)^{1/3} \\
& \approx 65 \mbox{ K} \left( \frac{k_{\rm cond}}{10^{10} \mbox{ erg K}^{-1} \mbox{ s}^{-1}} \right)^{1/3} \left( \frac{R}{R_\oplus} \right)^{-1/3}.
\end{split}
\label{eq:cond}
\end{equation}
The condition in equation (\ref{eq:cond}) implies that as long as the rocky surfaces of close-in super Earths are heated to $\gtrsim 100$ K, thermal conductivity will not operate rapidly enough to heat the cold nightside hemisphere.  The possibility of significant geothermal heating from the core is not considered in our models and remains a topic for future investigation.

\section{Stability Diagrams}
\label{sect:stability}

\subsection{Conditions Related to Advection \& Radiation}

Since the atmosphere is predominantly forced by stellar irradiation, the main effect is the competition between advection and radiative cooling.  A plausible, necessary condition for atmospheric stability is $t_{\rm adv} < t_{\rm rad}$, which yields
\begin{equation}
R < \left( \frac{c_P u_{\rm max} P}{\sigma_{\rm SB} g} \right) T_\star^{-3} R_\star^{-3/2} \left( 1 - {\cal A} \right)^{-3/4} a^{3/2}.
\label{eq:condition1}
\end{equation}
The maximum zonal wind speed, $u_{\rm max}$, is obtained from our shallow water model (see \S\ref{subsect:shallow}).  A reasonable approximation to the condition in equation (\ref{eq:condition1}) is to set $u_{\rm max} = c_s$, which yields
\begin{equation}
\begin{split}
R <& \left( \frac{\Gamma k_{\rm B}}{m} \right)^{1/4} \left( \frac{3 c_P P}{4 \pi G \rho_0 \sigma_{\rm SB} } \right)^{1/2} \\
&\times T_\star^{-5/4} R_\star^{-5/8} \left( 1 - {\cal A} \right)^{-5/16} a^{5/8}.
\end{split}
\label{eq:condition1b}
\end{equation}
Since free gravity waves have a speed $\sim c_s$, the condition in equation (\ref{eq:condition1b}) may also be interpreted as a constraint on heat redistribution mediated by free gravity waves.

We have made the assumption that if convective instability is triggered, it leads to the redistribution of heat primarily in the vertical and meridional directions (and not in the zonal direction), which is corroborated by three-dimensional simulations of atmospheric circulation for tidally-locked, Earth-like exoplanets \citep{j03,ms10,hv11,hfp11}.

\subsection{Conditions Related to Thermodynamics of Phase Changes}

If advection manages to redistribute heat across the entire exoplanetary atmosphere, then it may be described by an equilibrium temperature $T_{\rm eq} = T_{\rm irr}/\sqrt{2}$.  Thus, we may identify the temperature in equation (\ref{eq:psat}) as $T=T_{\rm eq}$.  

Generally, the atmosphere remains in the gas phase if
\begin{equation}
a < \mbox{max}\left\{ a_{\rm con}, a_{\rm crit} \right\}.
\end{equation}
The condensation distance is given by 
\begin{equation}
a_{\rm con} \equiv \frac{R_\star}{2} \left( \frac{T_\star}{T_{\rm con,0}} \right)^2 ~\left( 1 - {\cal A} \right)^{1/2} ~\left[ \ln{\left( \frac{P_{\rm sat,0} }{P}\right)} \right]^2,
\label{eq:acon}
\end{equation}
while the critical distance is given by
\begin{equation}
a_{\rm crit} \equiv \frac{R_\star}{2} \left( \frac{T_\star}{T_{\rm crit}} \right)^2 \left( 1 - {\cal A} \right)^{1/2}.
\label{eq:acrit}
\end{equation} 
Typically, we have $a_{\rm crit} < a_{\rm con}$ because $T_{\rm crit} > T_{\rm con}$.

In other words, an atmosphere consisting of a single chemical species (described by $T_{\rm con,0}$ and $P_{\rm sat,0}$) will remain stable if the exoplanet is located sufficiently close to its host star, such that the atmospheric molecule (or atom) remains in the gas phase.  If the atmosphere contains several chemical species, then $P$ refers to the partial pressure of a given species.  In the present study, we consider only atmospheres consisting predominantly of H$_2$ and N$_2$.

\subsection{Conditions Related to Orbital Circularization \\ \& Spin Synchronization} 

Our application of the \cite{sp11} shallow water model breaks down---and the concerns about atmospheric stability are alleviated or even obviated---if the exoplanet is not spin-synchronized, but still (approximately) applies if it resides on a mildly eccentric orbit.\footnote{We will see later that the $t_{\rm adv} = t_{\rm rad}$ and the $t_{\rm circ} = t_\star$ lines typically do not intersect.}  (See \citealt{laughlin09} for an illustration of the effects of stellar irradiation on the atmosphere of an exoplanet residing on a highly eccentric, but spin-synchronized, orbit.  See also \citealt{heller11}.)  The time scale associated with spin synchronization is \citep{boden01}
\begin{equation}
t_{\rm syn} = \frac{8Q}{45 \Omega} \left( \frac{\omega}{\Omega} \right) \left( \frac{M}{M_\star} \right) \left( \frac{a}{R} \right)^3,
\label{eq:syn}
\end{equation}
where $\omega$ is the \emph{rotational} frequency of an exoplanet which is not initially spin-synchronized.  The time scale associated with the circularization of the orbit is \citep{gs66}
\begin{equation}
t_{\rm circ} = \frac{4Q}{63 \Omega} \left( \frac{M}{M_\star} \right) \left( \frac{a}{R} \right)^5.
\label{eq:circ}
\end{equation}
In both equations (\ref{eq:syn}) and (\ref{eq:circ}), the tidal quality factor is $Q = 2\pi E_{\rm peak}/\Delta E$, where $E_{\rm peak}$ is the peak tidal energy stored and $\Delta E$ is the amount of energy dissipated per forcing cycle \citep{knopoff64,gs66,boden01}.  Following \cite{knopoff64} and \cite{gs66}, we adopt $Q=10$--100 for rocky exoplanets.  Spin synchronization generally occurs faster than orbital circularization, unless
\begin{equation}
a < R \sqrt{ \frac{14}{5} \left( \frac{\omega}{\Omega} \right) }.
\label{eq:circ_fast}
\end{equation}
If $\omega/\Omega = 100$, then the numerical coefficient in equation (\ref{eq:circ_fast}) is about 17.  This condition appears unlikely to be satisfied.  Demanding that $t_{\rm syn} < t_\star$ and $t_{\rm circ} < t_\star$ yield a pair of respective conditions:
\begin{equation}
\begin{split}
&a < \left( \frac{135 t_\star}{32 \pi Q \rho_0} \right)^{2/9} \left( \frac{\Omega}{\omega} \right)^{2/9} G^{1/9} M_\star^{1/3}, \\
&R > \left( \frac{16 \pi Q \rho_0}{189 t_\star} \right)^{1/2} G^{-1/4} M_\star^{-3/4} a^{13/4}. \\
\end{split}
\label{eq:tidal}
\end{equation}

A caveat is that if the exoplanet ends up in $1:N$ tidal resonances where $N>1$, then this tends to homogenize the zonal temperature differences, thus further alleviating concerns about atmospheric collapse \citep{wordsworth11}.

\subsection{Other Conditions}

In order to retain its atmosphere, a super Earth needs to possess a radius which is smaller than its Hill sphere $R_{\rm H} = a (M/3M_\star)^{1/3}$, which implies that
\begin{equation}
a > \left( \frac{9 M_\star}{4 \pi \rho_0} \right)^{1/3}.
\label{eq:hill}
\end{equation}
The thermal speed ($\sim c_s$) of the atmospheric gas also needs to not exceed the escape speed of the exoplanet, $v_{\rm esc} = ( 2 G M/R)^{1/2}$,
\begin{equation}
R > \left( \frac{3 \Gamma k_{\rm B} }{ 8 \pi G \rho_0 m} \right)^{2/3} T^{2/3}_\star \left( 1 - {\cal A} \right)^{1/6} a^{-1/3}.
\label{eq:escape}
\end{equation}
Equation (\ref{eq:escape}) may be regarded as being liberal, as atmospheric escape (e.g., \citealt{yelle04,mc09}) will still occur when $v_{\rm esc}>c_s$.  We do not expect the lower limit in equation (\ref{eq:escape}) to be revised upward by more than an order of magnitude if atmospheric escape is properly accounted for (and if it occurs at a sufficiently high rate to be relevant over the stellar age).

The habitable zone, which is the range of distances from a star where an Earth-like exoplanet can permanently harbor liquid water, may be described by fitting formulae based on one-dimensional radiative-convective calculations \citep{selsis07}:
\begin{equation}
\begin{split}
&a_{\rm in} = 1\mbox{ AU} ~\left( a_1 - a_2 \tilde{T}_\star - a_3 \tilde{T}_\star^2 \right) \left( \frac{{\cal L}_\star}{{\cal L}_\odot} \right)^{1/2},\\
&a_{\rm out} = 1\mbox{ AU} ~\left( a_4 - a_5 \tilde{T}_\star - a_6 \tilde{T}_\star^2 \right) \left( \frac{{\cal L}_\star}{{\cal L}_\odot} \right)^{1/2},
\end{split}
\label{eq:hz}
\end{equation}
where $\tilde{T}_\star \equiv T_\star - 5700$ K, ${\cal L}_\star = 4 \pi R_\star^2 \sigma_{\rm SB} T^4_\star$ and ${\cal L}_\odot = 3.839 \times 10^{33}$ erg s$^{-1}$.  The fitting coefficients are $a_1 = 0.84$, $a_2 = 2.7619 \times 10^{-5}$ K$^{-1}$, $a_3 = 3.8095 \times 10^{-9}$ K$^{-2}$, $a_4 = 1.67$, $a_5 = 1.3786 \times 10^{-4}$ K$^{-1}$ and $a_6 = 1.4286 \times 10^{-9}$ K$^{-2}$.  The quantities $a_{\rm in}$ and $a_{\rm out}$ represent the inner and outer boundaries, respectively.  For simplicity, we have adopted a cloud-free description of the habitable zone; in such a scenario, the habitable zone for the Solar System extends from about 0.84--1.67 AU.

\subsection{Basic Stability Diagram}

\begin{figure}
\centering
\includegraphics[width=\columnwidth]{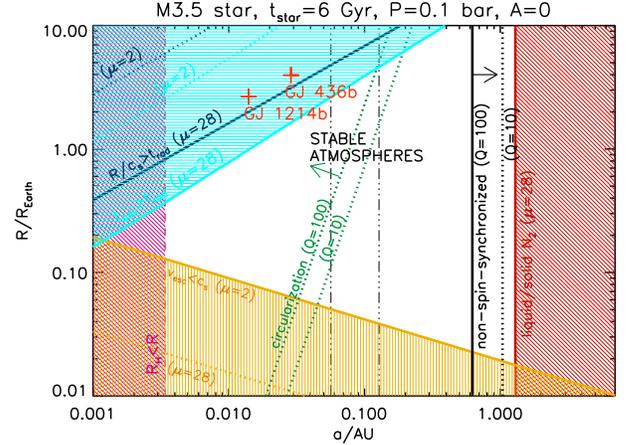}
\caption{Basic stability diagram in the parameter space of exoplanetary radius ($R$) versus the spatial separation between the exoplanet and its host star ($a$).  The hatched regions mark the regimes in which the atmosphere is unstable.  The pressure associated with the infrared photosphere is set to be $P=0.1$ bar, while the Bond albedo is ${\cal A}=0$.  The boundaries of the habitable zone are marked by the pair of vertical, triple-dot-dash lines.  For H$_2$, we have $a_{\rm con} \approx 24$ AU, and thus it is not shown in the diagram.}
\label{fig:stability1}
\end{figure}

Figure \ref{fig:stability1} shows the basic stability diagram for super Earth atmospheres in the parameter space of $R$ versus $a$.  For the condition $t_{\rm syn} < t_\star$, we have used $\omega/\Omega = 100$ merely as an illustration.  As expected, the condition $R/c_s < t_{\rm rad}$ is less constraining than $t_{\rm adv} < t_{\rm rad}$ across all of the considered values of $R$, $a$ and $\mu$; reassuringly, both conditions produce similar qualitative trends.  The condition $R_{\rm H} > R$ is overwhelmed by both the $t_{\rm adv} < t_{\rm rad}$ and $v_{\rm esc} > c_s$ conditions.  A key feature of Figure \ref{fig:stability1} is that atmospheres with lower mean molecular weights ($\mu$) tend to occupy a larger allowed region of parameter space.

We include in Figure \ref{fig:stability1} the measured values of $R$ and $a$ for GJ 436b and GJ 1214b.  For GJ 436b, we use the values of $R \approx 3.96 R_\oplus$ and $a \approx 0.0287$ AU as reported by \cite{knutson11}.  For GJ 1214b, we use the values of $R \approx 2.66 R_\oplus$ and $a \approx 0.014$ AU as reported by \cite{carter11}. Since both GJ 436b and GJ 1214b have a bulk density of $\rho_0 \approx 2$ g cm$^{-3}$, we adopt this value.  We adopt stellar parameters appropriate to a M3.5 star ($M_\star = 0.184 M_\odot$, $R_\star = 0.203 R_\odot$, $T_\star = 3240$ K) as a compromise between the M2.5 and M4.5 host stars of GJ 436b and GJ 1214b, respectively \citep{ed11}; we will refine these parameter values later in \S\ref{subsect:stability_specific}.  We also use $t_\star=6$ Gyr to be consistent with the reported ages of the host stars.

Another key feature of Figure \ref{fig:stability1} is that some super Earths ($R \gtrsim R_\oplus$) residing in the habitable zone (equation [\ref{eq:hz}]), with nitrogen-dominated (``Earth-like") atmospheres ($\mu=28$), do not possess stable atmospheres.  Certainly, this statement depends on the assumed value of the infrared photospheric pressure $P$.  Although we do not explicitly include it in our stability diagram, we have repeated the exercise for oxygen-dominated atmospheres ($\mu=32$) and find that the conclusions associated with them are similar to those for nitrogen-dominated atmospheres.  Thus, it is reasonable to term nitrogen-dominated atmospheres as being ``Earth-like".

\cite{kite11} have identified a pair of climate instabilities, related to weathering, which are suppressed for atmospheres with high radiative efficiencies (i.e., short $t_{\rm rad}$).  Specifically, the enhanced substellar weathering instability (ESWI) operates when advection is efficient enough to counteract the increase in temperature near the substellar point, which decreases the rate of weathering.  The supply of greenhouse gases (via outgassing) now outweighs its removal, implying that a runaway greenhouse effect may be triggered.  Thus, ESWI is activated when advection is efficient, but it acts on geological, rather than dynamical, time scales.  This has the implication that even if irradiation is strong enough to suppress advection and the ESWI, atmospheric collapse will still occur.  At face value, our results are orthogonal to those of \cite{kite11}, implying that super Earths which manage to overcome the ESWI need to also possess sufficiently advective atmospheres to avoid collapse.  Future work will elucidate the restricted parameter space in which both effects are precluded.

\subsection{Stability Diagrams Applied to GJ 436b and GJ 1214b}
\label{subsect:stability_specific}

To construct stability diagrams in the parameter space of $\mu$ versus $P$, we rewrite the condition in equation (\ref{eq:condition1}):
\begin{equation}
\mu < \left[ \frac{\left( 2 + n_{\rm dof} \right) {\cal R}^\ast u_{\rm max} P}{2 \sigma_{\rm SB} g R} \right] T_\star^{-3} R_\star^{-3/2} \left( 1 - {\cal A} \right)^{-3/4} a^{3/2}.
\label{eq:condition2}
\end{equation}

For GJ 436, we use $M_\star = 0.45 M_\odot$, $R_\star = 0.46 R_\odot$ and $T_\star = 3585$ K \citep{ed11}.  For GJ 1214, we use $M_\star = 0.15 M_\odot$, $R_\star = 0.21 R_\odot$ and $T_\star = 3026$ K \citep{ed11}.  The surface gravities of GJ 436b and GJ 1214b are approximately 13.0 and 9.0 m s$^{-2}$, respectively.

\begin{figure}
\centering
\includegraphics[width=\columnwidth]{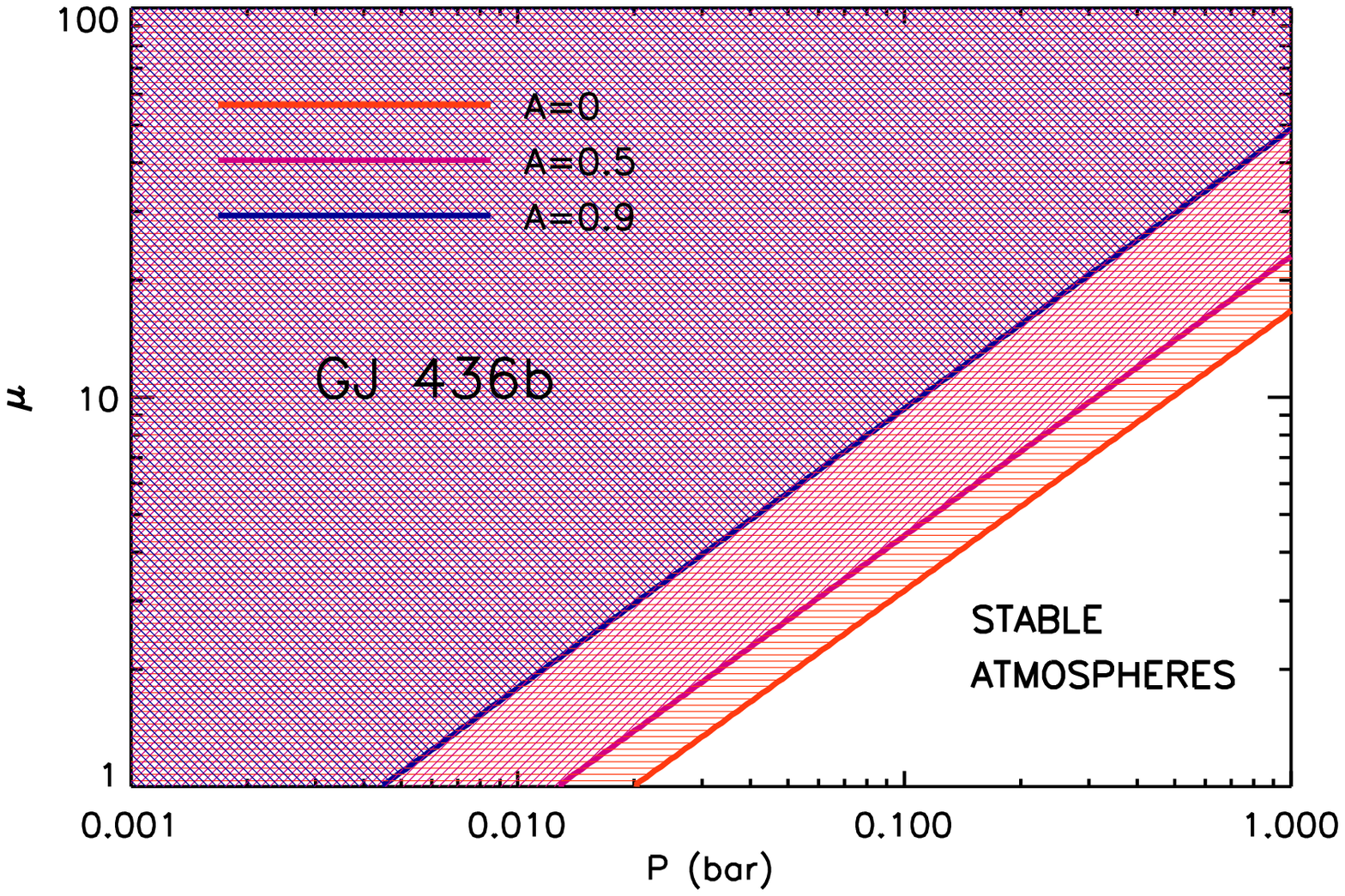}
\includegraphics[width=\columnwidth]{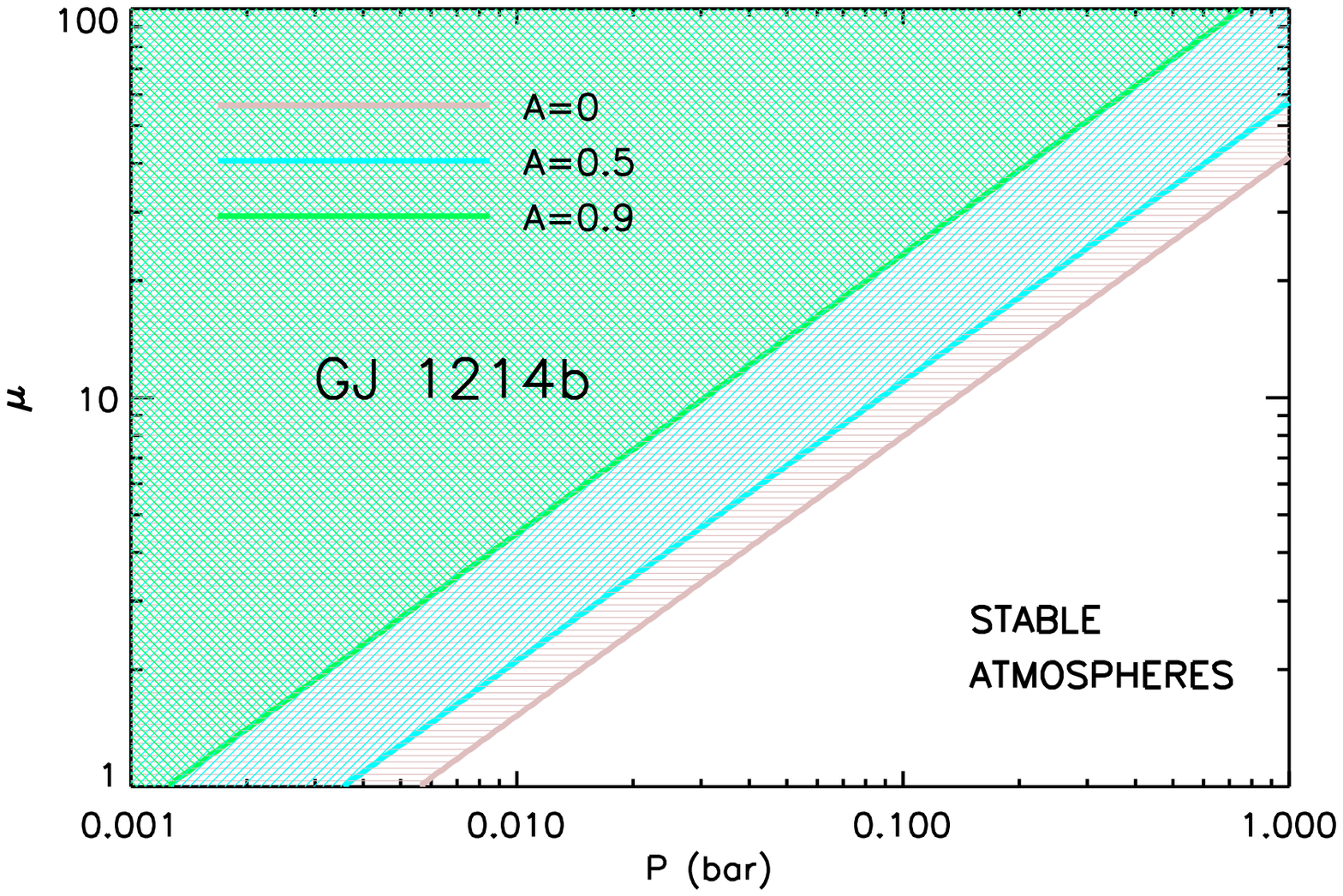}
\caption{Stability diagrams in the parameter space of mean molecular weight ($\mu$) versus infrared photospheric pressure ($P$), specialized to GJ 436b (top panel) and GJ 1214b (bottom panel).  The hatched regions mark the regimes in which the atmosphere is unstable for a given value of the Bond albedo (${\cal A}$).  The stellar ($M_\star$, $R_\star$ and $T_\star$) and exoplanetary ($R$, $a$) parameters are all set to their observed values for each exoplanet (see text).}
\label{fig:stability2}
\end{figure}

Figure \ref{fig:stability2} shows the stability diagrams as functions of $\mu$ and $P$, specialized to GJ 436b and GJ 1214b.  In general, atmospheres with high mean molecular weights ($\mu$) and low $P$ values are unlikely to be stable due to the violation of the $t_{\rm adv} < t_{\rm rad}$ condition. 

The \textit{Spitzer} transmission spectrum of GJ 436b reveals variations indicative of spectral absorption features associated with carbon monoxide and methane \citep{knutson11}.  Our stability diagram for GJ 436b allows for a stable atmosphere with $\mu \sim 10$ if the photosphere resides at $P \gtrsim 1$ bar and ${\cal A}=0$.  Non-zero values of the Bond albedo allow for higher values of $\mu$.  Inferences about the atmospheric composition of GJ 1214b remain controversial \citep{rs10,bean11,croll11,desert11,demooij12,mrk12}, including the possibility that a reduced value of the radius---which may obtain from an updated estimate of the distance to GJ 1214 via improved parallex measurements---is consistent with an atmosphere-less exoplanet.  Detecting or ruling out the presence of a Rayleigh scattering slope in the spectrum at shorter wavelengths is a valuable atmospheric diagnostic for $\mu$ \citep{demooij12}.

\section{Discussion}
\label{sect:discussion}

\begin{figure}
\centering
\includegraphics[width=\columnwidth]{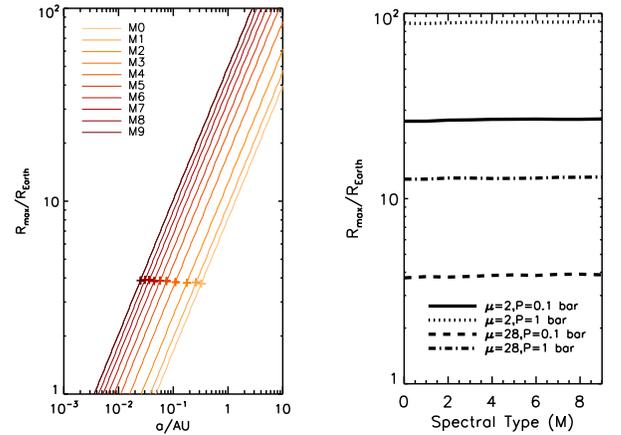}
\caption{Maximum exoplanetary radius $R_{\rm max}$ allowed by the $t_{\rm adv} < t_{\rm rad}$ condition across the spectral types M0 to M9.  The left panel shows the set of $R_{\rm max}$ curves, versus $a$, for $\mu=28$ and $P=0.1$ bar.  The crosses mark the location of the center of the habitable zone for each spectral type.  The right panel shows the $R_{\rm max}$ value corresponding to the center of each habitable zone as a function of the spectral type and for different combinations of the mean molecular weight $\mu$ and infrared photospheric pressure $P$.}
\label{fig:mstars}
\end{figure}

\begin{figure}
\centering
\includegraphics[width=\columnwidth]{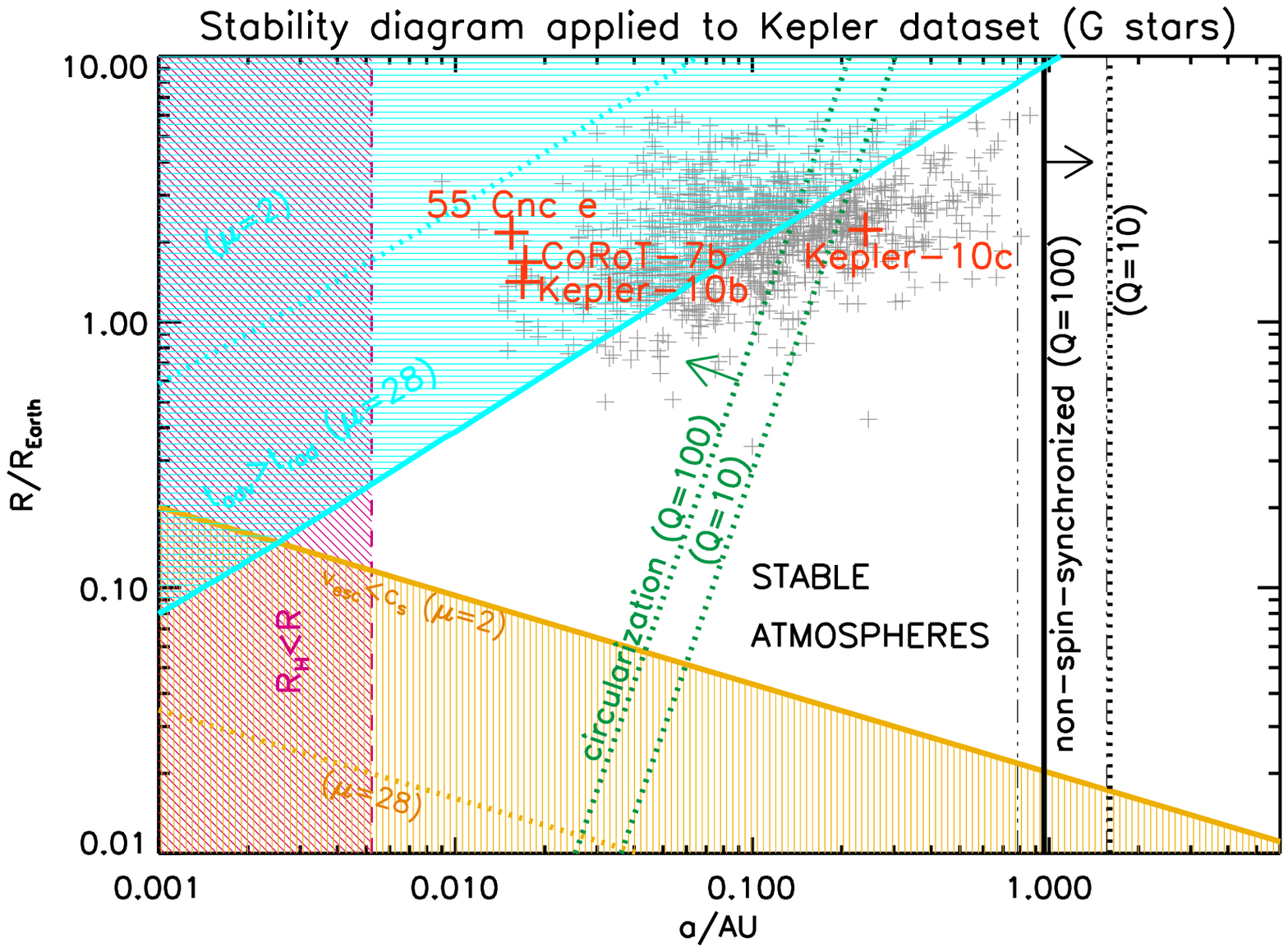}
\includegraphics[width=\columnwidth]{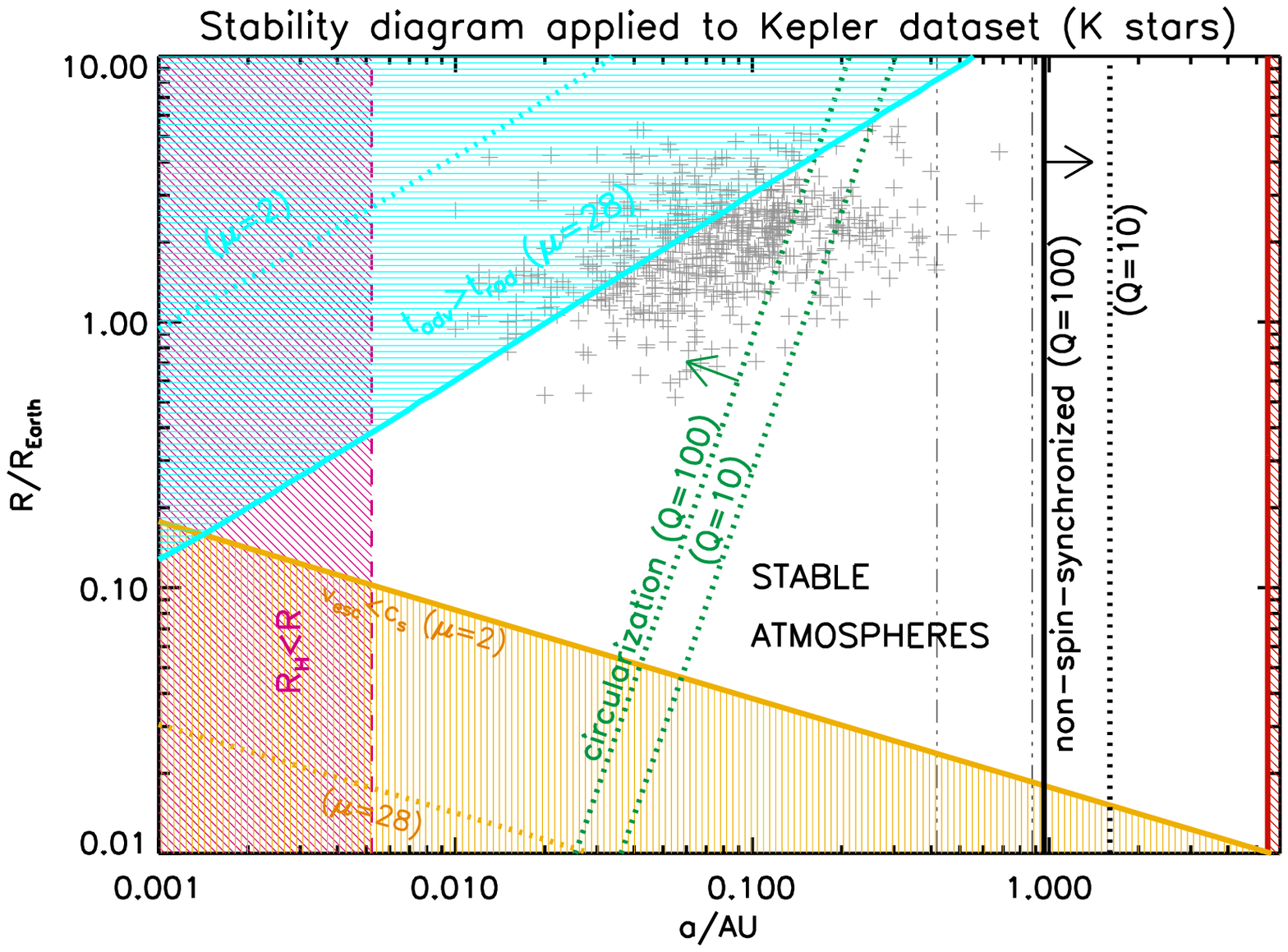}
\caption{Basic stability diagrams applied to the \textit{Kepler} dataset for G (top panel; 1135 objects) and K (bottom panel; 613 objects) stars.  The photospheric pressure is assumed to be $P = 1$ bar in these diagrams (see text).  The triple-dot-dash lines denote the boundaries of the habitable zones.  For the stability diagram applied to G stars, we include 55 Cancri e and CoRoT-7b as well as highlighting Kepler-10b and Kepler-10c.  For the top panel, we have $a_{\rm con} \approx 10$ and 136 AU for N$_2$ and H$_2$, respectively; for the bottom panel, these values are about 5 and 71 AU.}
\label{fig:kepler}
\end{figure}

The theory of (exo)planet formation remains fraught with uncertainties.  Therefore, an attempt to construct a formation model for super Earth atmospheres---which are probably secondary (i.e., due to outgassing) and unlikely to reflect the elemental compositions of the primordial nebulae from which they formed---remains an unconstrained exercise in some instances (e.g., \citealt{miguel11}).  From this perspective, our analytical framework may be regarded as a survival---rather than formation---model, similar to what has been constructed for planetesimal disks \citep{ht10}.

\subsection{Caveats, Future work \& Comparisons to Previous Work}
\label{subsect:caveats}

We have assumed that when advection is inefficient, the nightside of a tidally-locked super Earth becomes arbitrarily cold.  In reality, a non-negligible amount of heat may be advected to the nightside when $t_{\rm adv} \sim t_{\rm rad}$.  The temperature difference between the dayside and the nightside is \citep{sg02},
\begin{equation}
\Delta T \sim T_{\rm irr} \left[ 1 - \exp{\left(-t_{\rm adv}/t_{\rm rad} \right)} \right].
\end{equation}
Since $\Delta T \sim T_{\rm irr} - T_{\rm night}$, where $T_{\rm night}$ is the average temperature of the nightside hemisphere, we obtain
\begin{equation}
\frac{t_{\rm adv}}{t_{\rm rad}} \sim \ln{\left[ \frac{T_\star}{T_{\rm night}} \left( \frac{R_\star}{a} \right)^{1/2} \left( 1 - {\cal A} \right)^{1/4} \right]}
\end{equation}
such that $t_{\rm adv}/t_{\rm rad} \rightarrow \infty$ when $T_{\rm night} \rightarrow 0$; when $T_{\rm night} \rightarrow T_{\rm irr}$, we get $t_{\rm adv}/t_{\rm rad} \rightarrow 0$.
Adopting $T_\star = 3240$ K, $R_\star = 0.203 R_\odot$, $a=0.001$ AU and ${\cal A}=0$ yields $t_{\rm adv} \sim 6 t_{\rm rad}$ if $T_{\rm night} = 10$ K.  For $T_{\rm night} = 100$ K, we get $t_{\rm adv} \sim 3 t_{\rm rad}$.  In other words, chemical species with low condensation temperatures (such as H$_2$ with $T_{\rm con} \sim 10$ K) may remain in gaseous form even when the advective exceeds the radiative time scale by a factor of a few.  Specifically, it is possible for a trickle of heat to maintain an atmosphere if it is hydrogen-dominated even in cases where $t_{\rm adv} < t_{\rm rad}$ is unfulfilled.  Thus, the $t_{\rm adv} < t_{\rm rad}$ curve on the stability diagram should be viewed not as a sharp boundary, but a smooth transition.  In general, chemical species with very low condensation temperatures probably do not require $t_{\rm adv} < t_{\rm rad}$ as a necessary condition for atmospheric stability.  The gaseous state of these chemical species, with low condensation temperatures, may also be maintained by internal heat, which we do not consider.

Several other caveats provide motivation for future work.  We have only considered atmospheres consisting of a single chemical species, but this approximation should not be too severe in situations where the given gas (e.g., H$_2$, N$_2$) is the dominant constituent.  More refined calculations may consider cocktails of greenhouse gases.  Another assumption we have made is that our model, rocky exoplanets are not covered by oceans.  Oceanic and atmospheric dynamics couple in a non-trivial manner, over a broad range of time scales, with consequences for the heat redistribution between the day- and nightside hemispheres, and are beyond the scope of the present study.  Our hope is that our simple framework will motivate future work on a broad range of atmospheric constituents and not just Earth- or even Solar System-centric ones.  More refined trains of thought include the consideration of feedback mechanisms such as the water ice-albedo feedback, which has been shown to operate weakly on tidally-locked exoplanets with Earth-like stellar irradiation and atmospheric conditions \citep{j97,j03}.

Finally, we compare the present study to previous studies utilizing three-dimensional simulations of atmospheric circulation.  For example, the rocky, CO$_2$-dominated super Earth model considered by \cite{wordsworth11} has $t_{\rm rad} \sim 10^7$--$10^8$ s, but they do not report their computed zonal wind speeds and thus we are unable to determine $t_{\rm adv}$.  However, we can infer from their Figure 1 that even though their model, in 1:1 tidal resonance, appears to have $t_{\rm adv} \gtrsim t_{\rm rad}$, the nightside temperatures are about 280--290 K.  These temperatures lie above the condensation temperature of CO$_2$ at 20 bar, which is about 240 K, and thus atmospheric collapse does not occur.  \cite{j97} do not provide the value of $c_P$ used; if we use $c_P = 10^7$ erg K$^{-1}$ g$^{-1}$, then we obtain $t_{\rm rad} \sim 10^7$ s for their runs with $P \sim 1$ bar.  Since $u_{\rm max} \sim 10$ m s$^{-1}$, we have $t_{\rm adv} \sim 10^5$--$10^6$ s ($<t_{\rm rad}$) for their runs with $R = 6400$ and 13000 km, implying that these model atmospheres are mostly advective, which is consistent with the findings of \cite{j97}.

\subsection{Maximum Stability Radius}

Our basic stability diagram (Figure \ref{fig:stability1}) demonstrates that there is a maximum exoplanetary radius $R_{\rm max}$ above which the (thin) atmospheres of super Earths are unstable.  In the left panel of Figure \ref{fig:mstars}, we show $R_{\rm max}$ as a function of $a$ across the spectral types M0 to M9.  The less intensive irradiation from the cooler M stars leads to more lethargic radiative cooling in the atmospheres of their super Earths, which increases $R_{\rm max}$ across all values of $a$.  However, this effect is offset by the center of the habitable zone moving inwards, towards the star, for later spectral types.  The collective result is that $R_{\rm max}$ is insensitive to the spectral type.  Rather, it is somewhat sensitive to the mean molecular weight ($\mu$) and the infrared photospheric pressure ($P$), as shown in the right panel of Figure \ref{fig:mstars}.  Atmospheres with lower photospheric pressures and higher mean molecular weights have lower values of $R_{\rm max}$, implying that the condition for stability is more stringent.  For Earth-like atmospheres ($\mu=28$) and $g \sim 10$ m s$^{-2}$, the photospheres of super Earths with $R\sim 1$--10$R_\oplus$ need to be located at $P \gtrsim 0.1$ bar for stability.

\subsection{Application to Kepler Exoplanets and Exoplanet Candidates}

As a final demonstration of the power of our stability diagram, we apply it to the \textit{Kepler} dataset \citep{borucki11,batalha12}.  We include all exoplanets and exoplanet candidates with $R< 6R_\oplus$.\footnote{We remain agnostic about the \emph{theoretical} demarcation between Earth- and Neptune-like exoplanets.}  We construct two diagrams for G (5200 K $< T_\star \le 6000$ K) and K (3700 K $< T_\star \le 5200$ K) stars using the average values of the stellar parameters for each sub-sample.  We adopt $t_\star=5$ Gyr for illustration, noting that this only affects the conditions on circularization and spin synchronization.  In the absence of empirical constraints, we adopt Solar System-centric values for various parameters ($\rho_0 = 3$ g cm$^{-3}$, ${\cal A} = 0.3$), but note that our estimates are insensitive to these assumptions.  A key, unknown parameter is the infrared photospheric pressure.  In Figure \ref{fig:kepler}, we set $P=1$ bar.  With $P = 1$ bar, almost all of the 1135 \textit{Kepler} exoplanets and exoplanetary candidates orbiting G stars have stable atmospheres if the atmospheres are dominated by molecular hydrogen ($\mu=2$), whereas about 44\% of them have stable atmospheres if they are Earth-like in composition ($\mu=28$).  For $P = 0.1$ bar, these percentages become about 88\% and 1\%, respectively.

We include in the stability diagram for G stars the measured $R$ and $a$ values for 55 Cancri e ( $R \approx 2.17 R_\oplus$, $a \approx 0.01544$ AU; \citealt{demory11,gillon12}) and CoRoT-7b ($R \approx 1.68 R_\oplus$, $a \approx 0.0172$ AU; \citealt{leger09}).  We also highlight the exoplanets Kepler-10b ($R \approx 1.416 R_\oplus$, $a \approx 0.01684$ AU; \citealt{batalha11}) and Kepler-10c ($R \approx 2.227 R_\oplus$, $a \approx 0.2407$ AU; \citealt{fressin11}).  It remains possible that 55 Cancri e, CoRoT-7b and Kepler-10b may maintain minimal atmospheres---which do not extend beyond the day-night terminators and are dynamically dominated by vertical, rather than horizontal, flows---established through vapour saturation equilibrium with their continuously eroded rocky surfaces \citep{cm11,valencia11}.

For the K stars (613 exoplanets and exoplanet candidates), we obtain similar numbers: for $P = 0.1$ bar, we get 97\% ($\mu=2$) and 4\% ($\mu=28$); for $P = 1$ bar, we get 100\% ($\mu=2$) and 70\% ($\mu=28$).  Curiously, the exoplanet candidates and exoplanets residing within the habitable zones of both the \textit{Kepler} G and K stars are likely to possess stable atmospheres, unlike the situation with M stars.

If these exoplanetary atmospheres are Earth-like ($P = 1$ bar, $\mu = 28$), then about half of them are expected to be stable.  The continued expansion of the \textit{Kepler} dataset and the eventual, inexorable pursuit of follow-up observations to further characterize these exoplanets will allow our predictions to be tested.  For example, the photospheric infrared emission from atmosphere-less super Earths is likely to be emanated directly from their rocky surfaces, such that the dayside to nightside flux contrast will appear marked; these exoplanets will also exhibit flat transmission spectra.  The scrutiny of atmosphere-less super Earths may provide a survey of their surface compositions \citep{hu12}.

\acknowledgments
\textit{KH acknowledges generous support by the Zwicky Prize Fellowship of ETH Z\"{u}rich (Star and Planet Formation Group; PI: Michael Meyer).  We are grateful to Sascha Quanz, George Lake and Ren\'{e} Heller for illuminating conversations, as well as Jonathan Mitchell for stimulating comments which improved the clarity and quality of the manuscript.  We thank the anonymous referee for a thoughtful report which improved the robustness of the study.}


\begin{thebibliography}{}

\bibitem[Anglada-Escud\'{e} et al.(2012)]{ang12} Anglada-Escud\'{e}, G., et al. \ 2012, ApJ Letters, in press (arXiv:1202.0446)

\bibitem[Antuono(2010)]{ant10} Antuono, M. \ 2010, Journal of Fluid Mechanics, 658, 166

\bibitem[Batalha et al.(2011)]{batalha11} Batalha, N.M., et al. \ 2011, ApJ, 729, 27

\bibitem[Batalha et al.(2012)]{batalha12} Batalha, N.M., et al. \ 2012, preprint (arXiv:1202.5852) 

\bibitem[Bean et al.(2011)]{bean11} Bean, J.L., et al. \ 2011, ApJ, 743, 92

\bibitem[Bodenheimer, Lin \& Mardling(2001)]{boden01} Bodenheimer, P., Lin, D.N.C., \& Mardling, R.A. \ 2001, ApJ, 548, 466

\bibitem[Borucki et al.(2011)]{borucki11} Borucki, W.J., et al. \ 2011, ApJ, 736, 19

\bibitem[Carter et al.(2011)]{carter11} Carter, J.A., Winn, J.N., Holman, M.J., Fabrycky, D., Berta, Z.K., Burke, C.J., \& Nutzman, P. \ 2011, ApJ, 730, 82

\bibitem[Castan \& Menou(2011)]{cm11} Castan, T., \& Menou, K. \ 2011, ApJ, 743, L36

\bibitem[Charbonneau(2009)]{char09} Charbonneau, D. \ 2009, Proceedings of the IAU Symposium, eds. F. Pont, D. Sasselov \& M. Holman, 253, pg. 1--8

\bibitem[Charbonneau et al.(2009)]{char09b} Charbonneau, D., et al. \ 2009, Nature, 462, 891

\bibitem[Croll et al.(2011)]{croll11} Croll, B., Albert, L., Jayawardhana, R., Miller-Ricci Kempton, E., Fortney, J.J., Murray, N., \& Neilson, H. \ 2011, ApJ, 736, 78

\bibitem[de Mooij et al.(2012)]{demooij12} de Mooij, E.J.W., et al. \ 2012, A\&A, 538, A46

\bibitem[Delfosse et al.(2012)]{delfosse12} Delfosse, X., et al. \ 2012, preprint (arXiv:1202.2467)

\bibitem[Demory et al.(2011)]{demory11} Demory, B.-O., et al. \ 2011, A\&A, 533, 114

\bibitem[D\'{e}sert et al.(2011)]{desert11} D\'{e}sert, J.-M., et al. \ 2011, ApJ, 731, L40

\bibitem[Ehrenreich \& D\'{e}sert(2011)]{ed11} Ehrenreich, D., \& D\'{e}sert, J.-M. \ 2011, A\&A, 529, A136

\bibitem[Fressin et al.(2011)]{fressin11} Fressin, F., et al. \ 2011, ApJS, 197, 5

\bibitem[Frierson, Held \& Zurita-Gotor(2006)]{fhz06} Frierson, D.M.W., Held, I.M., \& Zurita-Gotor, P. \ 2006, Journal of the Atmospheric Sciences, 63, 2548

\bibitem[Garratt(1994)]{g94} Garratt, J.R. \ 1994, Earth-Science Reviews, 37, 89

\bibitem[Gill(1980)]{gill80} Gill, A.E. \ 1980, Quarterly Journal of the Royal Meteorological Society, 106, 447

\bibitem[Gillion et al.(2012)]{gillon12} Gillon, M., et al. \ 2012, A\&A, 539, A28

\bibitem[Goldreich \& Soter(1966)]{gs66} Goldreich, P., \& Soter, S. \ 1966, Icarus, 5, 375

\bibitem[Goody \& Yung(1989)]{gy89} Goody, R.M., \& Yung, Y.L. \ 1989, Atmospheric radiation: theoretical basis, 2nd edition (New York: Oxford University Press)

\bibitem[Held \& Suarez(1994)]{hs94} Held, I.M., \& Suarez, M.J. \ 1994, Bulletin of the American Meteorological Society, 75, 1825

\bibitem[Heller, Leconte \& Barnes(2011)]{heller11} Heller, R., Leconte, J., \& Barnes, R. \ 2011, A\&A, 528, A27

\bibitem[Heng \& Spitkovsky(2009)]{hs09} Heng, K., \& Spitkovsky, A. \ 2009, ApJ, 703, 1819

\bibitem[Heng \& Tremaine(2010)]{ht10} Heng, K., \& Tremaine, S. \ 2010, MNRAS, 401, 867

\bibitem[Heng, Menou \& Phillipps(2011)]{hmp11} Heng, K., Menou, K., \& Phillipps, P.J. \ 2011, MNRAS, 413, 2380

\bibitem[Heng \& Vogt(2011)]{hv11} Heng, K., \& Vogt, S.S. \ 2011, MNRAS, 415, 2145

\bibitem[Heng, Frierson \& Phillipps(2011)]{hfp11} Heng, K., Frierson, D.M.W., \& Phillipps, P.J. \ 2011, MNRAS, 418, 2669

\bibitem[Heng et al.(2012)]{hhps12} Heng, K., Hayek, W., Pont, F., \& Sing, D.K. \ 2012, MNRAS, 420, 20

\bibitem[Heng(2012)]{heng12} Heng, K. \ 2012, ApJ, 748, L17

\bibitem[Hu, Ehlmann \& Seager(2012)]{hu12} Hu, R., Ehlmann, B.L., \& Seager, S. \ 2012, ApJ, in press (arXiv:1204.1544) 

\bibitem[Joshi, Haberle \& Reynolds(1997)]{j97} Joshi, M.M., Haberle, R.M., \& Reynolds, R.T. \ 1997, Icarus, 129, 450

\bibitem[Joshi(2003)]{j03} Joshi, M. \ 2003, Astrobiology, 3, 415

\bibitem[Kite, Gaidos \& Manga(2011)]{kite11} Kite, E.S., Gaidos, E., \& Manga, M. \ 2011, ApJ, 743, 41

\bibitem[Knopoff(1964)]{knopoff64} Knopoff, L. \ 1964, Reviews of Geophysics, 2, 625

\bibitem[Knutson et al.(2011)]{knutson11} Knutson, H.A., et al. \ 2011, ApJ, 735, 27

\bibitem[Laughlin et al.(2009)]{laughlin09} Laughlin, G., Deming, D., Langton, J., Kasen, D., Vogt, S., Butler, P., Rivera, E., \& Meschiari, S. \ 2009, Nature, 457, 562

\bibitem[L\'{e}ger et al.(2009)]{leger09} L\'{e}ger, A., et al. \ 2009, A\&A, 506, 287

\bibitem[Longuet-Higgins(1968)]{lh68} Longuet-Higgins, M.S. \ 1968, Philosophical Transactions of the Royal Society of London, 262, 511

\bibitem[MacDonald, Lee \& Xie(2000)]{mac00} MacDonald, A.E., Lee, J.L., \& Xie, Y. \ 2000, Journal of the Atmospheric Sciences, 57, 2493

\bibitem[Matsuno(1966)]{matsuno66} Matsuno, T. \ 1966, Journal of the Meteorological Society of Japan, 44, 25

\bibitem[Mayor et al.(2009)]{mayor09} Mayor, M., et al. \ 2009, A\&A, 507, 487

\bibitem[Menou(2012)]{menou12} Menou, K. \ 2012, ApJ, 744, L16

\bibitem[Merlis \& Schneider(2010)]{ms10} Merlis, T.M., \& Schneider, T. \ 2010, Journal of Advances in Modeling Earth Systems --- Discussion (JAMES-D), 2, 13

\bibitem[Miguel et al.(2011)]{miguel11} Miguel, Y., Kaltenegger, L., Fegley, B., \& Schaefer, L. \ 2011, ApJ, 742, L19

\bibitem[Miller-Ricci Kempton, Zahnle \& Fortney(2012)]{mrk12} Miller-Ricci Kempton, E., Zahnle, K., \& Fortney, J.J. \ 2012, ApJ, 745, 3

\bibitem[Murray-Clay, Chiang \& Murray(2009)]{mc09} Murray-Clay, R.A., Chiang, E.I., \& Murray, N. \ 2009, ApJ, 693, 23

\bibitem[Perna, Heng \& Pont(2012)]{php12} Perna, R., Heng, K., \& Pont, F. \ 2012, ApJ, in press (arXiv:1201.5391v2)

\bibitem[Pierrehumbert(2010)]{pierrehumbert10} Pierrehumbert, R.T. \ 2010, Principles of Planetary Climate (New York: Cambridge University Press)

\bibitem[Pierrehumbert(2011)]{p11} Pierrehumbert, R.T. \ 2011, ApJ, 726, L8

\bibitem[Rogers \& Seager(2010)]{rs10} Rogers, L.A., \& Seager, S \ 2010, ApJ, 716, 1208

\bibitem[Selsis et al.(2007)]{selsis07} Selsis, F., Kasting, J.F., Levrard, B., Paillet, J., Ribas, I., \& Delfosse, X. \ 2007, A\&A, 476, 1373

\bibitem[Showman \& Guillot(2002)]{sg02} Showman, A.P., \& Guillot, T. \ 2002, A\&A, 385, 166

\bibitem[Showman \& Polvani(2010)]{sp10} Showman, A.P., \& Polvani, L.M. \ 2010, Geophysical Research Letters, 37, L18811

\bibitem[Showman \& Polvani(2011)]{sp11} Showman, A.P., \& Polvani, L.M. \ 2011, ApJ, 738, 71

\bibitem[Tarter et al.(2007)]{tarter07} Tarter, J., et al. \ 2007, Astrobiology, 7, 30

\bibitem[Troen \& Mahrt(1986)]{tm86} Troen, I.B., \& Mahrt, L. \ 1986, Boundary-Layer Meteorology, 37, 129

\bibitem[Wordsworth et al.(2011)]{wordsworth11} Wordsworth, R.D., Forget, F., Selsis, F., Millour, E., Charnay, B., \& Madeleine, J.-B. \ 2011, ApJ, 733, L48

\bibitem[Valencia et al.(2011)]{valencia11} Valencia, D., Ikoma, M., Guillot, T., \& Nettelmann, N. \ 2011, A\&A, 516, A20

\bibitem[Vallis(2006)]{vallis} Vallis, G.K. \ 2006, Atmospheric and Oceanic Fluid Dynamics: Fundamentals and Large-Scale Circulation (New York: Cambridge University Press)

\bibitem[Vogt et al.(2010)]{vogt10} Vogt, S.S., Butler, R.P., Rivera, E.J., Haghighipour, N., Henry, G.W., \& Williamson, M.H. \ 2010, ApJ, 723, 954

\bibitem[Yelle(2004)]{yelle04} Yelle, R.V. \ 2004, Icarus, 170, 167

\end{thebibliography}
\end{document}